\documentclass[aps,pra,notitlepage,reprint,onecolumn,showpacs,preprintnumbers,amsmath,amssymb,superscriptaddress,longbibliography,12pt]{revtex4-2}
\usepackage[section]{placeins}
\usepackage{graphicx}

\usepackage{amsmath}
\usepackage{amssymb}
\usepackage{contour}
\usepackage{dsfont}
\usepackage{bm}

\graphicspath{{compiled_figures/}{figures/}{}}

\usepackage{hyperref}
\hypersetup{pdfstartview={FitH},pdfpagemode={UseNone},
            colorlinks,linkcolor=blue, citecolor=blue, urlcolor=blue,
            bookmarksopen=true, pdfnewwindow=true}
\usepackage[all]{hypcap}

\newcommand{\subsectionJournal}[1]{{\bf #1.}}

\begin{document}
\sloppy
\title{Complex-birefringent dielectric metasurfaces\\ for arbitrary polarization-pair transformations}

\author{Shaun Lung}
\email{lung.shaun@gmail.com}
\affiliation{ARC Centre of Excellence for Transformative Meta-Optical Systems (TMOS), Nonlinear~Physics Centre, Research School of Physics,\\ The Australian National University, Canberra, ACT 2601, Australia}
\author{Kai Wang}
\affiliation{ARC Centre of Excellence for Transformative Meta-Optical Systems (TMOS), Nonlinear~Physics Centre, Research School of Physics,\\ The Australian National University, Canberra, ACT 2601, Australia}
\affiliation{Ginzton Laboratory and Department of Electrical Engineering,\\ Stanford University, Stanford, CA 94305, USA}
\author{Khosro Zangeneh Kamali}
\affiliation{ARC Centre of Excellence for Transformative Meta-Optical Systems (TMOS), Department of Electronic Materials Engineering, Research School of Physics,\\ The Australian National University, Canberra, ACT 2601, Australia}
\author{Jihua Zhang}
\affiliation{ARC Centre of Excellence for Transformative Meta-Optical Systems (TMOS), Nonlinear~Physics Centre, Research School of Physics,\\ The Australian National University, Canberra, ACT 2601, Australia}
\author{Mohsen~Rahmani}
\author{Dragomir~N.~Neshev}
\affiliation{ARC Centre of Excellence for Transformative Meta-Optical Systems (TMOS), Department of Electronic Materials Engineering, Research School of Physics,\\ The Australian National University, Canberra, ACT 2601, Australia}
\author{Andrey A. Sukhorukov}
\email{andrey.sukhorukov@anu.edu.au}
\affiliation{ARC Centre of Excellence for Transformative Meta-Optical Systems (TMOS), Nonlinear~Physics Centre, Research School of Physics,\\ The Australian National University, Canberra, ACT 2601, Australia}

\begin{abstract}
Birefringent materials or nanostructures that introduce phase differences between two linear polarizations underpin the operation of wave plates for polarization control of light. Here we develop metasurfaces realizing a distinct class of complex-birefringent wave plates, which combine polarization transformation with a judiciously tailored polarization-dependent phase retardance and amplitude filtering via diffraction. We prove that the presence of loss enables the mapping from any chosen generally non-orthogonal pair of polarizations to any other pair at the output. We establish an optimal theoretical design-framework based on pairwise nanoresonator structures and experimentally demonstrate unique properties of metasurfaces in the amplification of small polarization differences and polarization coupling with unconventional phase control. Furthermore, we reveal that these metasurfaces can perform arbitrary transformations of biphoton polarization-encoded quantum states, including the modification of the degree of entanglement. Thereby, such flat devices can facilitate novel types of multi-functional polarization optics for classical and quantum applications.
\end{abstract}

\maketitle
Nanostructured optical metasurfaces have appeared as an important class of modern photonic elements that can replace or outperform the capabilities of conventional optics~\cite{Yu:2014-139:NMAT}. In particular, recent advances in all-dielectric metasurfaces~\cite{Kuznetsov:2016-846:SCI} that do not suffer from plasmonic loss have led to demonstrations of very efficient transmissive optics. Importantly, such ultra-compact devices can realize flexible and highly precise manipulation and imaging of polarization states~\cite{Bomzon:2001-1424:OL, Arbabi:2015-937:NNANO, Pors:2015-716:OPT, Kruk:2017-2638:ACSP, Mueller:2017-113901:PRL, Wang:2018-1104:SCI, Stav:2018-01101:SCI, Neshev:2018-58:LSA, Shi:2020-eaba3367:SCA}, in particular providing a replacement for transitionally bulky wave plates~\cite{Kruk:2017-2638:ACSP} and performing polarization-spatial conversion for classical~\cite{Bomzon:2001-1424:OL, Arbabi:2015-937:NNANO, Pors:2015-716:OPT, Mueller:2017-113901:PRL} and quantum light~\cite{Wang:2018-1104:SCI, Stav:2018-01101:SCI, Neshev:2018-58:LSA, Georgi:2019-70:LSA}. Yet these applications are typically based on the realization of %real-valued 
phase retardance along the ordinary and extraordinary axes of nanoresonators, whereas the amplitudes of basis polarizations remain unchanged, making use of the inherently real-valued birefringence.

\begin{figure}[bt]
    \includegraphics[width=1\columnwidth]{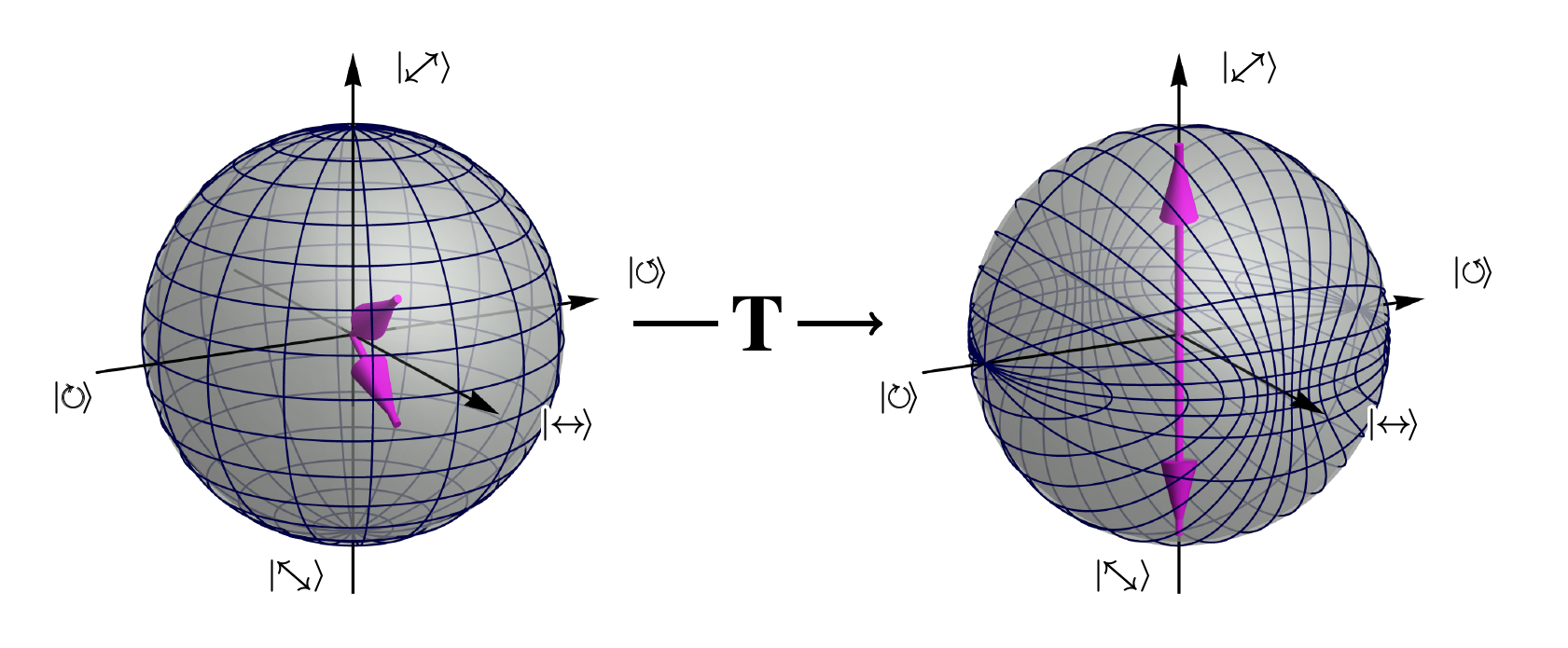} % tikzline
    \caption{Transformation of polarization states with a complex-birefringent wave plate, visualized on a Poincar\'e sphere. Left: two input states with close polarization state vectors indicated by purple arrows. Right: state vectors (purple arrows) after a transmission via a complex-birefringent wave plate, which become orthogonal. Gridlines on the surfaces illustrate the transformations for other states. 
    } % tikzline
    \label{fig:transPair}
\end{figure}

A new regime of the so-called \emph{complex} birefringence has recently been suggested~\cite{Cerjan:2017-253902:PRL}, extending the notion of real-valued birefringence. It was predicted that the presence of tailored polarization-dependent gain and loss can enable fundamentally new possibilities for polarization control, which cannot be achieved with conventionally birefringent media alone.
For example, complex birefringent structures could be used to change the angle between a pair of polarization states, as illustrated in Fig.~\ref{fig:transPair}. As shown, a pair of nearby polarization vectors can become orthogonal after propagating in the complex-birefringent medium, whereas in conventional birefringence the relative angle remains constant. Such a feature can deliver important benefits for a variety of applications, including improving the experimental measurement and detection responsivity.

More generally, complex birefringence realizes effective non-Hermitian Hamiltonian for the transformation of polarization states, which opens new regimes for the manipulation of quantum states and photon interference based on non-conservative systems~\cite{Roger:2015-7031:NCOM, Vest:2017-1373:SCI, Tischler:2018-21017:PRX}. Non-conservative optical transformations can also underpin the construction of optical neural networks~\cite{Shen:2017-441:NPHOT}.
However, the previously theoretically proposed scheme for complex birefringence that relies on propagation in a complex three-dimensional metamaterial incorporating subwavelength loss and gain materials~\cite{Cerjan:2017-253902:PRL} is extremely challenging and remains inaccessible for experimental implementations. Until now, a practical realization of the complex birefringence concept is still lacking. In addition, utilizing gain to accomplish complex birefringence 
essentially prevents the manipulation of 
quantum states. Indeed,  
the optical gain introduces and amplifies quantum noise, thus presenting an obstacle for quantum applications~\cite{Clerk:2010-1155:RMP}. 
On the other hand, material absorption can increase the temperature of the material and induce thermal instability on the refractive index and thus affect the optical properties of the system.

Dielectric metasurfaces have recently been demonstrated to be a versatile platform for the manipulation and measurement of both classical and quantum light~\cite{Stav:2018-01101:SCI, Wang:2018-1104:SCI, Georgi:2019-70:LSA}. Compared with three-dimensional metamaterials, single-layer metasurfaces are much easier to fabricate. 
In this work, we suggest and realize experimentally a new conceptual approach for implementing complex birefringence with all-dielectric metasurfaces, without material gain or absorption. 
We formulate a practical design principle that is optimal for achieving any desirable polarization transformation with a minimum amount of loss, which is realized through judiciously engineered polarization-dependent diffraction. 
Such loss can be effectively eliminated for certain polarization states, presenting a fundamental advantage compared to the 
material loss exhibited by all states in plasmonic structures Ref.~\cite{Vest:2017-1373:SCI}. Our approach can be tailored for arbitrary polarization pairs along with polarization transformation and phase control, which can underpin a new generation of polarization optical devices for classical and quantum applications.

\section*{\uppercase{Results and discussion}}
The concept of complex birefringence may be described mathematically by contrast to conventional birefringence. Consider a pair of states $|A_i\rangle$ and $|B_i\rangle$, where, for a unitary transformation associated with real birefringence, defining a target transformation $\textbf{T}|A_i\rangle = |A_t\rangle$ simultaneously restricts the transformation of the corresponding paired state $\textbf{T}|B_i\rangle = |B_t\rangle$, as 
$\langle A_t |B_t \rangle = \langle A_i|B_i\rangle$. This equality means that the relative angle between the vectors defining polarization states is preserved on the Poincar\'e sphere, and in particular a transformation sketched in Fig.~\ref{fig:transPair} cannot be achieved. Thus, while it is possible to realize the transformation of any pre-defined polarization state to another, arbitrary simultaneous transformations for pairs of states cannot be realized using conventional birefringence~\cite{Yu:2014-139:NMAT}.

\subsectionJournal{Theoretical approach}
Now we establish a theoretical approach to realize complex birefringence in metasurfaces. Mathematically, the target transformation of optical waves is
defined by a non-unitary transfer matrix $\textbf{T}$ of dimension $2\times2$, which acts on the input polarization state vector to determine the output polarization state. For this purpose, we apply the singular value decomposition (SVD), which has been used previously to design non-conservative optical transformations in waveguide circuits~\cite{Miller:2012-23985:OE, Shen:2017-441:NPHOT, Tischler:2018-21017:PRX}. Here, we apply SVD in a different context for designing transmission through an ultra-thin metasurface, and represent the desired transfer matrix as
\begin{align} \label{eq:decomposition_SVD}
    \textbf{T} &= 
        \textbf{U}
        \begin{bmatrix}
            \sigma_1 & 0 \\
            0 & \sigma_2
        \end{bmatrix}
        \textbf{W}^\dag 
        ,
\end{align}
where $\textbf{U}=[\textbf{U}_1,\textbf{U}_2]$ and $\textbf{W}=[\textbf{W}_1,\textbf{W}_2]$ are unitary matrices which subscripts refer to columns, and $\sigma_{1,2}$ are the non-negative singular values obtained through SVD. To be specific, we assume that $\sigma_1 \ge \sigma_2$. This decomposition reveals the physical meaning of the complex birefringence. The input polarization state $\textbf{W}_1$ is transformed to $\sigma_1 \textbf{U}_1$ at the output. In absence of gain, this state is fully transmitted when $\sigma_1 = 1$. The orthogonal input polarization state, $\textbf{W}_2$, is attenuated and transformed into an output state $\sigma_2 \textbf{U}_2$. Thereby, the transformation $ \textbf{T}$ incorporates simultaneously a wave-plate %{\color{blue} (There is no phase retardance difference after the transformation $\sigma_{1,2}$ are real.)} 
and a selective polarization attenuator. It can also be interpreted as a general elliptical dichroism~\cite{Zhukovsky:2009-1988:OL, Plum:2009-113902:PRL, Hu:2017-41893:SRP} combined with tailored phase retardance of polarization eigenstates.
\begin{figure}[tb]
    \includegraphics[keepaspectratio=true,width=1\columnwidth,height=0.7\textheight]{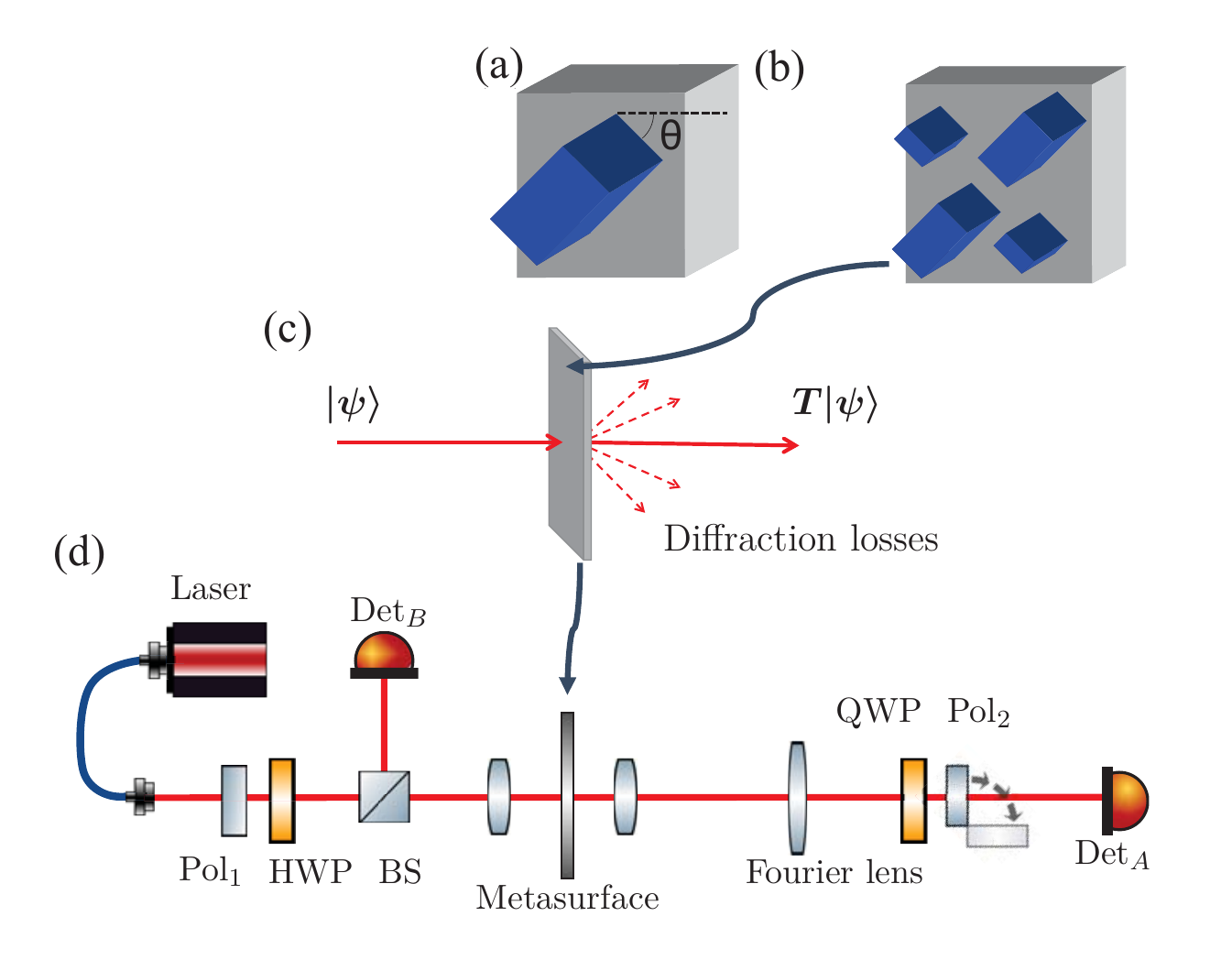} % tikzline
    \caption{ % tikzline
    Concept of realizing complex birefringence with all-dielectric metasurfaces. %(a)~Refractive index ellipse illustrating conventional polarization control comprising real-valued phases along the ordinary and extraordinary axes. 
    (a)~A single-structure unit cell with dielectric nano-pillar placed at an angle $\theta$ realizing real birefringence, with real-valued phase retardation along the two principal axes. (b)~A unit cell with two pairs of distinctly sized nano-pillars realizing complex birefringence. (c)~A metasurface utilizing the unit cell in (b), realizing engineered polarization-dependent losses via diffraction. (d)~Schematic diagram of the experimental setup utilized to characterize the manufactured samples. % tikzline
    } % tikzline
    \label{fig:binary_metasurface}
\end{figure}

\FloatBarrier

Next, we rearrange the terms in Eq.~(\ref{eq:decomposition_SVD}) as
\begin{align} \label{eq:decomposition_SVD2}
    \textbf{T} &=     
        \frac{\sigma_1}{2} \textbf{U}
        \begin{bmatrix}
            1 & 0 \\
            0 & e^{i \kappa}
        \end{bmatrix}
        \textbf{W}^\dag 
        + 
        \frac{\sigma_1}{2} \textbf{U}
        \begin{bmatrix}
            1 & 0 \\
            0 & e^{-i \kappa}
        \end{bmatrix}
        \textbf{W}^\dag
        = \frac{\sigma_1}{2} (\textbf{T}_1 + \textbf{T}_2) ,
\end{align}
where $\kappa = \cos^{-1}(\sigma_2 / \sigma_1)$. By construction, $\textbf{T}_1$ and $\textbf{T}_2$ are unitary matrices. 
This decomposition reveals a possibility of performing arbitrary complex transformation with binary metasurfaces, composed of two types of resonators, each realizing the unitary transformation $\textbf{T}_1$ and $\textbf{T}_2$. 
We note that this decomposition is applicable to arbitrary transfer matrices, where in general all elements of $\textbf{T}$ may be different. 

\subsectionJournal{Dielectric metasurface design}
For the experimental demonstration, we consider metasurfaces composed of non-chiral nanoresonators, since chiral response is usually weak in all-dielectric structures.
Remarkably, we find that arbitrary 
transformation from a pair of generally non-orthogonal polarizations to another polarization-pair state can be achieved 
by single-layer all-dielectric metasurfaces. We note that at normal incidence, the transfer matrix has to be symmetric with equal values of off-diagonal elements~\cite{Shi:2020-eaba3367:SCA}. We find an explicit solution for such transfer matrix as follows:
\begin{equation} \label{eq:TsymmAB}
    \textbf{T} = \frac{e^{i\phi_g}}{\sigma_m} \textbf{T}_0 = \frac{e^{i\phi_g}}{\sigma_m} \left[\langle A_t^\ast|B_i\rangle \cdot |B_t\rangle\langle A_i^\perp| - \langle B_t^\ast|A_i\rangle \cdot |A_t\rangle\langle B_i^\perp| \right],
\end{equation}
where the subscripts $i$ and $t$ correspond to the input and transformed after the metasurface states, respectively, 
$\sigma_m$ is the maximum singular value of $\textbf{T}_0$, and $\phi_g$ is a global phase we can arbitrarily choose. The vectors $A_i^\perp=i\varsigma_2 A_i^\ast$ and $B_i^\perp=i\varsigma_2 B_i^\ast$ are the orthogonal states of $A_i$ and $B_i$, respectively, where $\varsigma_2$ is the second Pauli matrix. This form of transfer matrix assures the $A_t$ and $B_t$ output states for the $A_i$ and $B_i$ input states, respectively. It is also a symmetric matrix as it can be checked by direct substitution that off-diagonal elements are equal to each other, $\langle H | \textbf{T} | V \rangle \equiv \langle V | \textbf{T} | H \rangle$.
We note that only some specific output states cannot be reached according to the form of Eq.~(\ref{eq:TsymmAB}), $\langle A_t^\ast|B_i\rangle \ne 0$ and $\langle B_t^\ast|A_i\rangle \ne 0$. While such a constraint could be removed by engineering non-symmetric transfer matrices if required, we focus in the following on a range of practical applications based on the symmetric transfer matrix form.
We now outline how to implement transformations governed by any symmetric matrix $\textbf{T}$.
In this regime, the $\textbf{T}_{1,2}$ defined in Eq.~(\ref{eq:decomposition_SVD2}) are unitary symmetric matrices, and can be represented~\cite{Arbabi:2015-937:NNANO} as
\begin{align}
    \textbf{T}_{k}
    &=  \textbf{R}(\theta^{(k)})
        \begin{bmatrix}
            \textrm{e}^{i\phi_o^{(k)}} & 0 \\
            0 & \textrm{e}^{i\phi_e^{(k)}}
        \end{bmatrix}
        \textbf{R}(-\theta^{(k)}).
   \label{eq:decomposition_symmetrical}
\end{align}
Here the matrix $\textbf{R}(\theta)$ performs a rotation by angle $\theta$. Accordingly, Eq.~(\ref{eq:decomposition_symmetrical}) describes the polarization transformation by a single nano-resonator oriented at angle $\theta^{(k)}$ as illustrated in Fig.~\ref{fig:binary_metasurface}(a), where $\phi_o^{(k)}$ and $\phi_e^{(k)}$ are the phase accumulations along the two axes. This provides a direct path to the physical construction of a metasurface.

\FloatBarrier

According to the analytically established decomposition principle, we arrange the two types of nanoresonators in a unit cell of a binary metasurface, as shown in Fig.~\ref{fig:binary_metasurface}(b). 
We note that metasurfaces containing pairwise structures have been extensively explored for various applications such as circular microwave polarizers~\cite{Plum:2009-113902:PRL}, terahertz polarization gratings~\cite{Yang2018}, and optical holograms~\cite{Deng2018}, however their capacity for the realization of complex birefringence was not known.
The binary arrangement leads to the interference of the %differing 
optical polarization states transmitted through the individual nano-pillars in the forward direction of the metasurface, 
as illustrated in Fig.~\ref{fig:binary_metasurface}(c). This effectively implements the sum on the right-hand-side of Eq.~(\ref{eq:decomposition_SVD2}), delivering the desired complex transformation. In general, the transmission is accompanied by diffraction, which acts as a polarization-dependent loss channel. While such loss is inherently necessary for implementing complex birefringence, our all-dielectric metasurface design achieves the mathematically minimum required amount of loss without any material absorption. Specifically, in the optimal case of $\sigma_1=1$, according to Eq.~(\ref{eq:decomposition_SVD}) there exists a polarization state $|\textbf{W}_1\rangle$ which is fully transmitted with zero loss.

We perform detailed numerical modeling to design the binary metasurfaces with various transmission characteristics. The structures were tailored for operation at the wavelength of  $\lambda_0 = 1550\mathrm{nm}$, within the range of the telecommunications band. The nanoresonators are placed on a square lattice. We then selected the 
structure period
as $d=1800\mathrm{nm}$, which satisfies the following requirement: \textbf{(i)}~$d > \lambda_0$ to allow diffraction, which acts as polarization-dependent loss, and \textbf{(ii)}~nano-resonators are positioned sufficiently close to suppress direct light transmission through the substrate. %{\color{blue}do we want to avoid near field coupling?)}.
We consider resonators in the form of rectangular cuboids made of amorphous-silicon on a glass substrate, and achieve control of transmission phase along the extraordinary and ordinary axes ($\phi_{e}$ and $\phi_{o}$) by modifying their physical dimensions ($L_e$, $L_o$) in the plane of the metasurface~\cite{Arbabi:2015-937:NNANO}. These dimensions are determined by numerical simulations (see Materials and Methods).

\subsectionJournal{Experimental characterization}
The fabricated metasurfaces (see Materials and Methods) were characterized using the experimental setup shown in Fig.~\ref{fig:binary_metasurface}(d). A variable-wavelength laser operating in the $1500-1575\mathrm{nm}$ telecommunication range was used as the light source, and input polarization states were prepared from this laser using a fixed linear polarizer (Pol$_1$) and a motorized half-wave plate (HWP).
A 50:50 beam splitter (BS) and detector were introduced immediately after the state preparation to provide a normalization baseline for the measurements. A lens focused the beam to an approximate $80\mathrm{\mu m}$ spot normally incident on the metasurface, followed by an objective lens used to image the transmitted light. To exclusively collect zero-th order transmitted light, a Fourier lens was used to exclude higher orders of diffracted light.
Then, the polarization was characterized via a motorized quarter-wave plate (QWP) and a linear polarizer (Pol$_2$) to project the transmitted state onto a varying basis for reconstruction, as per
\begin{equation}
    P_{\rm Det_A} 
    = |\langle \theta_{\rm Pol_2}|\textbf{Q}(\theta_{Q})\textbf{T}|\theta_{in}\rangle|^2 P_{\rm Det_B} ,
    \label{eq:projection}
\end{equation}
where $\theta_{\rm Pol_2}$ is the orientation angle of the linear polarizer (Pol$_2$), $|\theta_{in}\rangle$ is the input linear polarization state at an angle of $\theta_{in}$ selected by a corresponding rotation of HWP, $\textbf{Q}(\theta_{Q})$ is the transmission matrix of the quarter-waveplate rotated at an angle $\theta_{Q}$, and $P_{\rm Det_{A,B}}$ are the measured powers at the detectors $A$ and $B$. %{\color{blue} (Eq. (\ref{eq:projection}) is actually the transmittance not the power reading on detector A)}. 
Readings were recorded over varying input states ($\theta_{in}$) and the angle of the QWP ($\theta_{Q}$) %{\color{blue} ($\theta_{Q}$ is the QWP angle, not the polarizer angle)} 
and subsequently numerically fitted to Eq.~(\ref{eq:projection}), allowing for the reconstruction of the symmetric transfer matrix \textbf{T}, up to a global phase.
Additionally, measurements were carried out with the final polarizer (Pol$_2$) removed to directly measure the total transmitted power without numerical fitting, and confirm the accuracy of the reconstructed transfer matrix.

\begin{figure}[tb]

    \includegraphics[width=1\columnwidth]{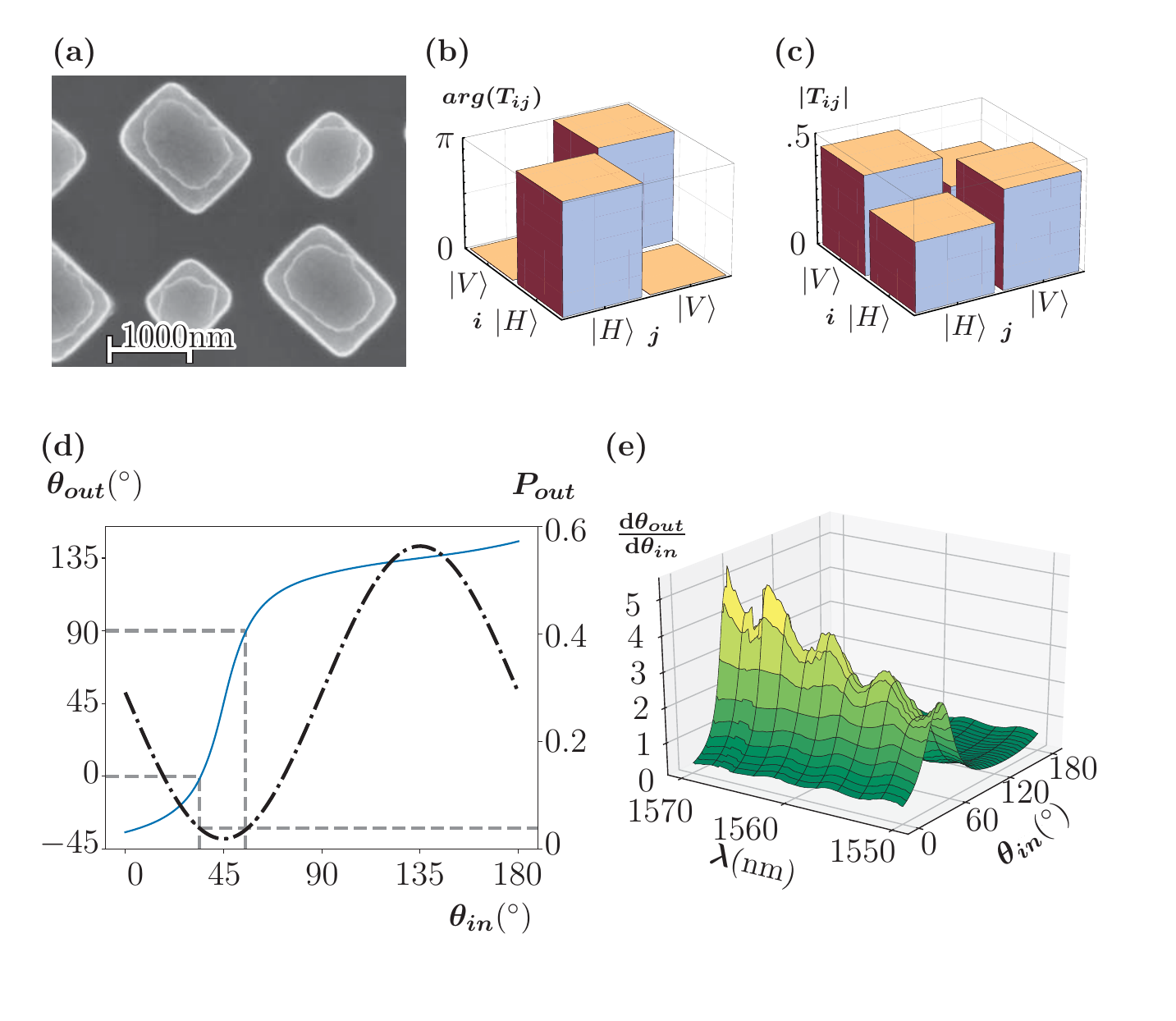} % tikzline
    \caption{Experimental results of a metasurface designed to bring nearby states into orthogonal ones. (a)~Scanning Electron Microscope (SEM) image of the manufactured metasurface. % tikzline
    (b,c)~Experimentally characterized arguments (b)~and absolute values (c) of the polarization transfer matrix \textbf{T} of the metasurface at a wavelength of $1563\mathrm{nm}$. (d)~Experimentally derived angular resolution of input linear polarization states denoted by $|\theta_{in}\rangle$ to the output states $|\theta_{out}\rangle$ and net transmitted power $P_{out}$, calculated at $1563\mathrm{nm}$ (scaled against input power of $0.995\mathrm{mW}$). Marked by dashed lines are the corresponding two states denoted by the arrows in Fig.~\ref{fig:transPair}.
    (e)~Responsivity ${\mathrm{d}\theta_{out}}/{\mathrm{d}\theta_{in}}$ of the metasurface vs. the input wavelength and polarization angle.} % tikzline
    \label{fig:metasurface_resolving}
    \label{fig:results_SEM_resolve} % tikzline
    \label{fig:results_phase_resolve} % tikzline
    \label{fig:results_magnitude_resolve} % tikzline
    \label{fig:trans_3D_resolving} % tikzline
    \label{fig:angles_resolving} % tikzline
\end{figure}

\subsectionJournal{Amplification of polarization differences}
We present two representative examples of fabricated metasurfaces with different functionalities. First,
we show the experimental realization of a metasurface that can transform a pair of very close polarizations into orthogonally polarized states. Specifically, we aim to demonstrate a transformation sketched in Fig.~\ref{fig:transPair} from the input linear polarization states $|A_i\rangle = |41^\circ\rangle$ and $|B_i\rangle = |49^\circ\rangle$
to the output states $|A_t\rangle = |0^\circ\rangle$ and $|B_t\rangle = |90^\circ\rangle$. %[Fig.~\ref{fig:transPair}(right)]. 
First, we apply Eq.~(\ref{eq:TsymmAB}) to determine the required transfer matrix.
Then, we employ the general theoretical framework to design a metasurface which can perform such transformation. 
In Fig.~\ref{fig:results_SEM_resolve}(a), we show the scanning electron microscopy (SEM) image of the fabricated metasurface, see Materials and Methods for details.
We measured the optical transmission through the metasurface using the setup outlined in Fig.~\ref{fig:binary_metasurface}(d) as described above, and reconstructed via tomography the experimental parameters of the metasurface transfer matrix, as shown in Figs.~\ref{fig:results_phase_resolve}(b,c) for the wavelength of $1563\mathrm{nm}$. It realizes a transformation of the nearby input states to the near-orthogonal output states $|A_t\rangle = |0.3\pm0.7^\circ\rangle$ and $|B_t\rangle = |89.8\pm0.7^\circ\rangle$, according to the theoretical design.

\FloatBarrier

We also analyze the singular value decomposition of the transfer matrix to quantify its efficiency. The singular values of the transfer matrix are found as $\sigma_1=0.751$ and $\sigma_2 = 0.138$, corresponding to maximally and minimally transmitted linear polarization states $|\textbf{W}_1\rangle = |135^\circ\rangle$ and $|\textbf{W}_2\rangle = |45^\circ\rangle$.
We note that the value of $\sigma_1$ below unity undesirably restricts the overall transmission, which can be increased through the improvements in sample fabrication and anti-reflection coating of the substrate. Yet most importantly, a small  ratio value of $\sigma_2/\sigma_1$ is an essential feature of a non-conservative transformation effected by complex birefringence. Indeed, under the Stokes formalism of polarization states, it is equivalent to the distortion and re-projection of states on the surface of the Poincar\'e sphere in a nonuniform manner. This allows one to overcome the fundamental limitations of real-valued birefringence that could only provide a global rotation of states across the Poincar\'e sphere, without changing the angle between pairs of polarization states. 
We confirmed experimentally that the metasurface can map nearby polarization states into well resolved orthogonal states, according to the concept in Fig.~\ref{fig:transPair}. We show in Fig.~\ref{fig:angles_resolving}(d) with a solid blue line a dependence of the output vs. input linear polarization angle. For the input states around $|\textbf{W}_2\rangle = |45^\circ\rangle$, there is an enhanced responsivity of the output polarization with respect to small changes of the input state, since ${\mathrm{d}\theta_{out}}/{\mathrm{d}\theta_{in}} \simeq 3.7 > 1$. In particular, the dashed lines mark two close input polarizations which are mapped to orthogonal states at the output.
Conversely, at input polarization angles around $|\textbf{W}_1\rangle = |135^\circ\rangle$, the metasurface provides the opposite effect, bringing well-separated polarization states closer together. We note that the increased responsivity is associated with a minimum of power transmission [dashed-dotted line in Fig.~\ref{fig:angles_resolving}(d)], which is a necessary tradeoff for the implementation of such functionality in all-dielectric structures without amplification.

We present the experimentally determined responsivity of mapping linear polarization states across a range of wavelengths in Fig.~\ref{fig:angles_resolving}(e). The visible oscillations vs. wavelength with a period of about $3nm$
are consistent with the Fabry-Perot interference due to a $170\mathrm{\mu m}$ thick glass substrate. Importantly, the increased responsivity is consistently present around the input angle $|\textbf{W}_2\rangle = |45^\circ\rangle$, for a wavelength range $1550-1570\mathrm{nm}$. The operating bandwidth can be further enhanced by specially optimizing the nano-resonator design and introducing anti-reflection coating on the substrate.

\begin{figure}[tb]

    \includegraphics[keepaspectratio=true,width=1\columnwidth,height=0.7\textheight]{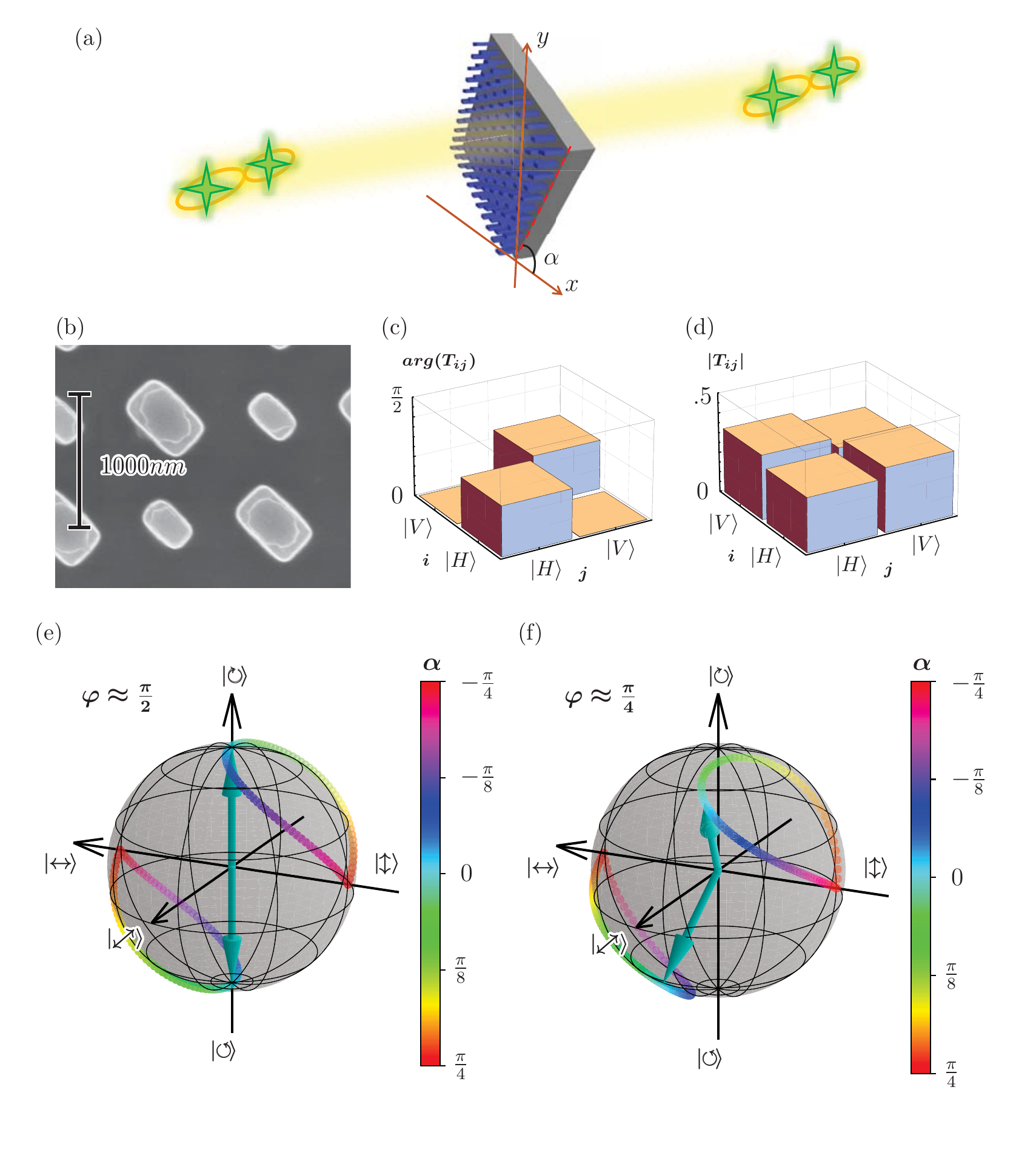} % tikzline
    \caption{ % tikzline 
    Application of complex-birefringent metasurfaces for arbitrary transformation of polarization-entangled two photon states. (a) Schematic of a metasurface oriented at an angle $\alpha$ to perform the desired transformation for entangled photon pairs. % tikzline
    (b)~SEM image of a metasurface made of amorphous-silicon on a glass substrate. (c,d)~Experimentally characterized transfer matrix at a wavelength of $1560\mathrm{nm}$, including (c)~arguments and (d)~modulus of the elements, which is close to Eq.~(\ref{eq:T_varphi}) with $\varphi=\pi/4$. % tikzline
    (e,f)~Transformation of $|\Psi(H,V)\rangle$ entangled  input state (e)~by an ideal phase retarder rotated at angle $\alpha$ and (f)~calculated with the experimental transfer matrix of the metasurface oriented at angle $\alpha$ as shown in~(a). % tikzline
    } % tikzline
    \label{fig:metasurface_nonhermitian}
    \label{fig:results_SEM_pi/4} % tikzline
    \label{fig:results_phase_pi/4} % tikzline
    \label{fig:results_magnitude_pi-4} % tikzline
    \label{fig:trans_slice_pi/4} % tikzline
    \label{fig:interference_pi/4} % tikzline
\end{figure}

\subsectionJournal{Transformation of quantum biphoton states}
Next, we reveal the capability for arbitrary transformation of quantum biphoton polarization states with complex birefringent metasurfaces, as sketched in Fig.~\ref{fig:metasurface_nonhermitian}(a). We note that any pure state of two polarization-entangled photons can be expressed as~\cite{Burlakov:2002-432:JETPL}
\begin{equation}
    |\Psi(A,B) \rangle = \frac{a^{\dag}(A) a^{\dag}(B) |0\rangle}{
        ||a^{\dag}(A) a^{\dag}(B)|0\rangle|| } ,
\end{equation}
where $A$ and $B$ are two single-photon polarization states, and $a^{\dag}$ is a one-photon creation operator. This expression means that a quantum two-photon state can be represented by a pair of points on the Poincar\'e sphere, for example such as those indicated by two arrows in Fig.~\ref{fig:transPair}(a).

The action of the metasurface on a two-photon state will, in general, produce a superposition of pure two-photon and mixed single-photon (when a paired photon is lost) states, which can be distinguished through conditional detection schemes. The two-photon state transformation is simply expressed as:
\begin{equation}
    |\Psi(A_i,B_i) \rangle \xrightarrow{\textbf{T}} |\Psi(\textbf{T} A_i, \textbf{T}B_i) \rangle = |\Psi(A_t, B_t) \rangle .
\end{equation}
This means that we can apply the approaches formulated above for the simultaneous independent transformation of a pair of polarization states according to Eq.~(\ref{eq:TsymmAB}). Importantly, the capacity of a complex birefringent metasurface to change the angle between the pair of states on a Poincar\'e sphere, such as illustrated in Fig.~\ref{fig:transPair}, enables the modification of the degree of quantum entanglement, which would be impossible with conventional conservative birefringence. Thereby, a transformation between arbitrary selected input to output quantum two-photon states can be accomplished.

Next, we showcase an experimental potential for tailored two-photon manipulation and quantum state transformation. Recent studies demonstrated that photon interference in lossy couplers can demonstrate unconventional features, including a transition between bunching and anti-bunching statistics~\cite{Vest:2017-1373:SCI}. The previous experiments were based on plasmonic structures with inherently high metal absorption. Here, we present a new design of all-dielectric metasurface which realizes a non-conservative coupler with the minimum necessary amount of loss. Specifically, we consider the coupling transformation between the $|H\rangle$ and $|V\rangle$ polarizations with a non-conservative transfer matrix~\cite{Vest:2017-1373:SCI} in the form:
\begin{equation}
    \label{eq:T_varphi}
    \textbf{T}_\varphi = \rho
    \begin{bmatrix}
        1 & \mathrm{e}^{i \varphi}\\
        \mathrm{e}^{i \varphi} & 1
    \end{bmatrix} ,
\end{equation}
where $\rho$ is a scaling coefficient. The phase value of  $\varphi=\pi/2$ corresponds to a conventional conservative coupler, while other phases generally correspond to non-conservative transformations. 
The optimal practical realization with the minimal necessary amount of losses in structures without gain corresponds to the maximum possible value of $\rho$, which is achieved when the maximum singular value of the transfer matrix $\sigma_1 \rightarrow 1$.

The general approach to complex birefringence formulated above enables us to design the binary metasurfaces which can realize the unconventional polarization coupling according to Eq.~(\ref{eq:T_varphi}) for arbitrary phase $\varphi$. We present an example of fabricated metasurface in Fig.~\ref{fig:metasurface_nonhermitian}(b), which transfer matrix [Figs.~\ref{fig:metasurface_nonhermitian}(c) and~(d)] realizes the case of $\varphi \simeq \pi/4$. There is only a slight variation of the absolute values in the experimental transfer matrix, which still enables unconventional interference and control of the biphoton states.

We use the experimentally determined transfer matrix [Fig.~\ref{fig:metasurface_nonhermitian}(c) and~(d)] to simulate the transformation of the input quantum state composed of orthogonally polarized entangled photons, $|\Psi(H,V) \rangle$, that is $A_i=H, B_i=V$. We further note that a rotation of a metasurface in the plane by angle $\alpha$, as sketched in Fig.~\ref{fig:metasurface_nonhermitian}(a),  nontrivially modifies the transformation as $\textbf{R}(\alpha) \textbf{T}_\varphi \textbf{R}(-\alpha)$ and thus the output states $A_t$ and $B_t$. We visualize the output two-photon state by a pair of same-color points on the Poincar\'e sphere in Fig.~\ref{fig:metasurface_nonhermitian}(f), where the colors correspond to different metasurface orientation angles $\alpha$ and the arrows point to the output state at $\alpha=0$. For comparison, we plot in Fig.~\ref{fig:metasurface_nonhermitian}(e) the output states which would be produced by a conventional conservative coupler with $\varphi=\pi/2$. We observe that in the conservative regime, the relative angle between the photon pairs is unchanged for all $\alpha$ [Fig.~\ref{fig:metasurface_nonhermitian}(e)], meaning that the degree of quantum entanglement at the output remains the same as at the input, in agreement with the properties of unitary operators. In contrast, this limitation is removed in the regime of complex birefringence, and the tailored change of relative angle [Fig.~\ref{fig:metasurface_nonhermitian}(f)] enables flexible entanglement control.
\FloatBarrier

\section*{\uppercase{Conclusion}}

In conclusion, we have formulated a provably optimal and practical design of all-dielectric metasurfaces realizing complex birefringent transformations, where only a minimally required amount of loss is introduced through diffraction without any material absorption. We have shown that even metasurfaces without a chiral response can realize mapping from any pair of polarization states to any output pair, which can enable arbitrary transformations of biphoton quantum states. We have fabricated and demonstrated experimentally metasurfaces for amplifying small differences in polarizations and realizing polarization couplers with unconventional phase control. The efficient operation of such ultra-thin meta-devices based on the principles of non-Hermitian optics can facilitate novel types of polarization state manipulation and measurement in both classical and quantum regimes.

\section*{\uppercase{Methods}}

\subsectionJournal{Fabrication of metasurfaces} 
The nanostructures were fabricated on
a 710-nm-thick amorphous-silicon thin film  prepared using Plasma-Enhanced Chemical Vapor Deposition (PECVD) on a $170\mathrm{\mu m}$ thick glass substrate. Then, etching was carried out via Electron Beam Lithography (EBL) and Inductively Coupled Plasma (ICP) etching. Varying exposures (from approximately $150\mathrm{\mu Ccm^{-2}}$ to $200\mathrm{\mu Ccm^{-2}}$) were utilized to produce a small range of metasurfaces with slight variations in sizes to account for fabrication variances.

It is worth noting that the shadows, which can be seen on the top of bars in the SEM images, Figs.~\ref{fig:metasurface_resolving}(a) and~\ref{fig:metasurface_nonhermitian}(b), are due to the coated E-spacer on the sample. E-spacer is generally used to make the sample conductive for electron microscopy purposes. But due to the significant difference in the thickness of our bars (710~nm), and E-spacer layer (20~nm), the coating cannot be done uniformly. Therefore those shadows are not avoidable. However, after taking SEM images, E-spacer was washed away by DI-water and did not exist during the optical measurements.

\subsectionJournal{Numerical simulations} 
The cross-section dimension parameters of individual nano-resonators were simulated using Rigorous Coupled Wave Analysis (RCWA)~\cite{Lalanne:1997-1592:JOSA, Lalanne:1996-779:JOSA}. Pairwise resonators were then arranged in a metasurface according to the form of Eq.~(\ref{eq:decomposition_symmetrical}) for a target transfer matrix, and the structure was optimized using a commercial electrodynamic solver CST Studio to fine-tune the parameters according to the phase-invariant fidelity measure 
$    \delta = 1 - {
        \left|\sum_{i,j}T^{\ast}_{ij} \widetilde{T}_{ij}\right|^2
    }\,{\left[
        \sum_{i,j}T^{\ast}_{ij}T_{ij}
        \sum_{i,j}\widetilde{T}^{\ast}_{ij}\widetilde{T}_{ij}\right]^{-1}
    },
    \label{eq:fidelity}$
where $T_{ij}$ and $\widetilde{T}_{ij}$ are the elements of the target and the numerically calculated transfer matrices, respectively. 
\section*{\uppercase{Acknowledgments}}
This work was supported by the Australian Research Council (DP160100619, DP190101559) and US AOARD (19IOA053). The metasurface fabrication was performed at The~Australian National University node of the Australian National Fabrication Facility (ANFF), a company established under the National Collaborative Research Infrastructure Strategy to provide nano and micro-fabrication facilities for Australia’s researchers.

   % bibliography data in db_Binary-MS.bib

\end{document}